\documentclass[12pt]{article}
\usepackage[english]{babel}
\usepackage {amssymb,amsmath,latexsym,enumerate,graphicx}
\usepackage{indentfirst} 
\usepackage{titlesec}
\usepackage {subfigure,epsfig}
\usepackage{caption2} 
\usepackage{cite}
\makeatletter

\renewcommand{\@biblabel}[1]{#1.} \makeatother

\makeatletter
\renewcommand{\section}{\@startsection%
{section}%
{1}%
{17pt}%
{-3.5ex plus -1ex minus -.2ex}%
{-10pt}%
{\textbf}%
} \makeatother

\begin{document}

%

\null
  \vskip 1.5em %
  \begin{center} %
       {\begin{tabular} [t]{c} %
        \large \lineskip .5em \textbf{Mikhail A. Chmykhov \ and \ Nikolai A. Kudryashov}
        \end{tabular} \par %
} \vskip 1.5em %
     {\textbf{APPROXIMATE SOLUTIONS OF NONLINEAR HEAT EQUATION FOR GIVEN FLOW\footnote{in Proceedings of
      the XXXIII Summer School Advanced Problems in Mechanics APM'2005, St.Petersburg, 2005}} \par} %
      \vskip 1.5em
     {Department of Applied Mathematics,Moscow  Engineering and Physics Institute (State university), 31
       Kashirskoe Shosse, 115409 Moscow, Russian Federation}
     {{\it E-mail:} chmykhov@mephi.ru, kudryashov@mephi.ru }
     \end{center} %
  \par
  \vskip 1.5em


{\footnotesize
\begin{quotation}
The one-dimensional problem of the nonlinear heat equation is
considered. We assume that the heat flow in the origin of coordinates
is the power function of time and the initial temperature is
zero. Approximate solutions of the problem are given. Convergence of
approximate solutions is discussed.
\end{quotation}
}

\section{The problem statement.}

The nonlinear heat equation has the form \cite{r1}

\begin{equation}\label{Eq_main}
\frac{\partial u}{\partial t} = {\varkappa} \frac{\partial}{\partial
r} \left( u^n \frac{\partial u}{\partial r} \right), \qquad r>0,
\qquad t>0
\end{equation}

where $ u(r,t) $ is temperature in the point $ r $; $ t $ is
time; $ \varkappa $ is coefficient of the heat conductivity.

Suppose the boundary condition has the form

\begin{equation}\label{Gr_u}
\left. u^n \frac{\partial u}{\partial r} \right|_{\theta = 0} = - q_0 t^k, \qquad t>0
\end{equation}

Where $ q_0 $ is constant. Equation (\ref{Gr_u}) corresponds to the energy flow in the origin of coordinates.

We also take the initial condition in the form

\begin{equation}\label{U_0}
u(r,t=0) = 0, \qquad r>0
\end{equation}

The boundary value problem (\ref{Eq_main})---(\ref{U_0}) was
considered \cite{r1, r2, r3, r4}. When the temperature specified on
the boundary the problem for equation (\ref{Eq_main}) were
considered in \cite{r3, r4, r5, r6, r7}. Numerical solution of the
problem (\ref{Eq_main})---(\ref{U_0}) can be found using difference
method \cite{7}.

In this work we are going to look for approximate solutions of the problem
(\ref{Eq_main})---(\ref{U_0}).

Using

\begin{equation*}\label{Zamena}
u = v^{\frac{1}{n}}
\end{equation*}

we get

\begin{equation}\label{Eq_new}
v_t = \varkappa v v_{r r} + \frac{\varkappa}{n} v_r^2, \qquad r>0,
\qquad t>0
\end{equation}

Taking the problem (\ref{Eq_main})---(\ref{U_0}) into consideration we
have the boundary and the initial conditions for equation (\ref{Eq_new})
in the form

\begin{equation}\label{Gr_Usl_new}
\left. v^{\frac{1}{n}} \frac{\partial v}{\partial r} \right|_{\theta
= 0} = -q_0 n t^k, \qquad t>0
\end{equation}

\begin{equation}\label{N_Usl_new}
v(r,t=0) = 0, \qquad r>0
\end{equation}

To solve the problem (\ref{Eq_new}) --- (\ref{N_Usl_new}) we can use
the self-similar variables \cite{r1, r2, r3, r4, r5, r6, r7, 8}

\begin{equation}\label{S-S_var}
v(r,t)=A t^m f(\theta), \qquad \theta = \frac{B r}{t^p}
\end{equation}

where $f(\theta )$ is a function of $ \theta $; $ A $, $ B $, $m$ and $p$ are constants.

Assuming

\begin{equation}\label{Param}
p = \frac{m+1}{2}, \qquad \varkappa A B^2 = 1
\end{equation}

and substituting (\ref{S-S_var}) into equation (\ref{Eq_new}) we obtain

\begin{equation}\label{ObshEq}
f f_{ \theta \theta } + {\frac {1}{n} f_{\theta}^2} +{\frac {m+1}{2}
\theta f_{\theta}} - mf  =0
\end{equation}

Substituting (\ref{S-S_var}) into the boundary condition
(\ref{Gr_Usl_new}) and taking into account

\begin{equation}\label{Param11}
\frac{m(n+1)}{n} - \frac{(m+1)}{2} = k
\end{equation}

\begin{equation}\label{Param12}
q_0 B^{-1} n = A^{\frac{n+1}{n}}
\end{equation}

we have the boundary condition for equation (\ref{ObshEq}) in the form

\begin{equation}
\left. f^{\frac{1}{n}} \frac{d f}{d \theta} \right|_{\theta = 0} =
-1
\end{equation}

From equation (\ref{N_Usl_new}) we find the second boundary condition

\begin{equation*}
f(\theta \to \infty) = 0
\end{equation*}

Using conditions (\ref{Param}), (\ref{Param11}) and (\ref{Param12})
we get constants $ A, \ B, \ m $ in form

\begin{equation}
A = \varkappa^{\frac{n}{n+2}} q_0^{\frac{2 n}{n+2}} n^{\frac{2
n}{n+2}}, \qquad B = \varkappa^{- \frac{n+1}{n+2}} q_0^{-
\frac{n}{n+2}} n^{- \frac{n}{n+2}}
\end{equation}


\begin{equation}\label{m0k}
m = \frac{n (2k+1)}{n+2}
\end{equation}

Taking expressions (\ref{Param}), (\ref{Param11}) and (\ref{Param12})
into account we have the boundary value problem
(\ref{Eq_main})---(\ref{U_0}) in the form

\begin{equation}\label{ObshEq_nu}
f f_{ \theta \theta } + {\frac {1}{n} f_{\theta}^2} +{\frac {m+1}{2}
\theta f_{\theta}} - m f =0
\end{equation}

\begin{equation}\label{Gr_lim}
\left. f^{\frac{1}{n}} \frac{d f}{d \theta} \right|_{\theta =0} = -1
\end{equation}

\begin{equation}\label{Gr_inf}
f(\theta \to \infty) = 0
\end{equation}

To find the solution of the problem (\ref{Eq_main}) --- (\ref{U_0}) we have to solve the problem
(\ref{ObshEq_nu}) --- (\ref{Gr_inf}). This is the aim of this work.

\section{Method applied.}

It is known that the velocity of the boundary for the nonlinear heat
conductivity is finite. Let us assume that $ \theta = \alpha $ is
the boundary. Therefore, at the point $ \alpha $ temperature is
equal to zero $( \ f(\alpha) = 0 \ )$, but derivative is non-zero
$\left( \
 \dfrac{{df}} {{d\theta}} \ne 0 \ \right) $.

We look for approximate solution of the problem (\ref{ObshEq_nu}) ---
(\ref{Gr_inf}) in the form

\begin{equation}\label{Ser_01}
f(\theta ) = \sum\limits_{i = 0}^N {\beta _i \left( {\alpha - \theta
} \right)^i }
\end{equation}

From equation (\ref{Ser_01}) we get

\begin{equation}\label{Proizvodnii}
\left. {\frac{{df}} {{d\theta }}} \right|_{\theta  = \alpha } = -
\beta _1 , \quad \left. {\frac{{d^2 f}} {{d\theta ^2 }}}
\right|_{\theta = \alpha }  = 2\beta _2 , \quad \ldots \quad, \left.
{\frac{{d^i f}} {{d\theta ^i }}} \right|_{\theta = \alpha } =
i!\left( { - 1} \right)^i \beta _i .
\end{equation}

Taking $ f(\alpha ) = 0 $ into account we get $ \beta_{{0}}=0 $

Substituting (\ref{Proizvodnii}) into equation (\ref{ObshEq}) we have the coefficients $
\beta_i $.

\begin{equation}
\begin{gathered}\label{BetaO_p1}
\beta_1 = \frac{1}{2} \alpha n \left( m + 1 \right)  \\
\beta_2 = \frac{1}{4} {\frac { \left( m - 1 \right) n}{n+1}} \\
\beta_3 = - \frac{1}{12} {\frac {n (m-1) \left( nm+n+2m \right)}{
\left( n+1 \right) ^{2}\alpha\, \left( 1+2\,n
\right) \left( m+1 \right) }} \\
\beta_4 = \frac{1}{48} {\frac {n (m-1) \left( nm+n+2m \right)
P_4^{(m,n)}}{{\alpha}^{2} \left( n+1 \right) ^{3} \left( 1+2\,n
\right)  \left( m+1 \right) ^{2 } \left( 3\,n+1 \right) }} \\
\beta_5 = - \frac{1}{240} {\frac {n (m-1) \left( nm+n+2m \right)
P_5^{(m,n)}}{{\alpha}^{3} \left( 3\,n+1 \right) \left( m+1 \right)
^{3} \left( 1+2\,n \right) ^{2} \left( n+1 \right) ^{4} \left(
1+4\,n \right) }}
\end{gathered}
\end{equation}

where $ P_4^{(m,n)} $ and $ P_5^{(m,n)} $ are polynomials

\begin{equation*}
\begin{gathered}
P_4^{(m,n)} = 5\,nm-n+7\,m-3 \\
\end{gathered}
\end{equation*}

\begin{equation*}
\begin{gathered}
P_5^{(m,n)} =
303\,n{m}^{2}+82\,{m}^{2}+102\,{n}^{3}{m}^{2}+317\,{n}^{2}{m}^{2}-204
\,{n}^{2}m \\
-238\,nm -70\,m-48\,{n}^{3}m+12+7\,{n}^{2}-6\,{n}^{3}+31\,n
\end{gathered}
\end{equation*}

Using expression (\ref{BetaO_p1}) we can obtain some exact solutions
of the problem (\ref{ObshEq_nu})
--- (\ref{Gr_inf}).

If $ n + 2 m + n m = 0 $ then $ \beta_i = 0 \quad (i \geq 3 $).
Taking $ m = - \frac{n}{n+2} $ into consideration we have exact
solution

\begin{equation}\label{Resh_0}
f(\theta) = \frac{1}{2} \alpha n \left( m + 1 \right) \left( {\alpha
- \theta } \right) + \frac{1}{4} {\frac { \left( m - 1 \right)
n}{n+1}} \left( {\alpha - \theta } \right)^2
\end{equation}

However this exact solution does not satisfy the boundary condition
(\ref{Gr_lim}). This solution can be used to solve the Cauchy
problem for equation (\ref{Eq_main}).

If $ m = 1 $ then $ \beta_i = 0 \quad (i \geq 2 $) and we have another exact solution for
equation (\ref{ObshEq_nu}).

\begin{equation}\label{Eq_2_12}
f(\theta) = \alpha n \left( {\alpha - \theta } \right)
\end{equation}

Exact solution (\ref{Eq_2_12}) satisfies the boundary condition
(\ref{Gr_lim}).

\section{Solutions of the problem (\ref{ObshEq_nu}) --- (\ref{Gr_inf}).}

Consider the case $ n > 0 $ and $ k = {1}/{n} $, ($ m = 1 $).

\begin{equation}\label{Solv_pl_1n}
f(\theta) = \alpha n \left( \alpha - \theta \right)
\end{equation}

Substituting (\ref{Solv_pl_1n}) into the boundary condition (\ref{Gr_lim}) we obtain

\begin{equation}\label{Find_a}
\frac{{\left( n+1 \right)} {\left( \alpha^2 n
\right)}^{\frac{n+1}{n}}}{n \alpha}=1
\end{equation}

From condition (\ref{Find_a}) we get the parameter  $ \alpha $

\begin{equation}
\alpha = \left( n^{\frac{1}{n}}(n+1) \right)^{\frac{n}{n+2}}
\end{equation}

The values of $ \alpha $ and $ n $ at $ k={1}/{n} $ are given in table \ref{t:pl}.

\begin{table}[h]
    \center
    \caption{} \label{t:pl}
    \begin{tabular}{||c|c|c|c|c|c|c||}     
        \hline
        n & 1 & $ {4}/{3} $ & $ 2 $ & $ {5}/{2} $ & 3 & 4\\ \hline
        $ \alpha $ & 1.2599 & 1.5299 & 2.0598 & 2.4586 & 2.8619 & 3.6840 \\
        \hline \hline
        n & $ {9}/{2} $ & $ 5 $ & $ {11}/{2} $ & 6 & $ {13}/{2} $ & $ 7 $ \\ \hline
        $ \alpha $ & 4.1025 & 4.5256 & 4.9528 & 5.3838 & 5.8184 & 6.2561  \\ \hline
    \end{tabular}
\end{table}

Exact solutions of the boundary value problem (\ref{ObshEq_nu}) --
(\ref{Gr_inf}) at $ k = {1}/{n} $ ($ n > 0 $) are described by
formula (\ref{Solv_pl_1n}).

Consider the case $ {n = 1} $ and $ k > 0 $. Approximate solution of
the boundary value problem (\ref{ObshEq_nu}) -- (\ref{Gr_inf}) can
be written in the form

\begin{equation}\label{f3_10}
\begin{gathered}
f(\theta) = \frac{1}{2} \alpha  \left( m + 1 \right)  \left(
\alpha-\theta \right) + \frac{1}{8} {\left( m - 1 \right) } \left(
\alpha-\theta \right) ^{2 } \\
- \frac{1}{144} {\frac { \left( 3\,m+1 \right)  \left( m-1 \right)
}{ \left( m+1 \right) \alpha}} {\left( \alpha-\theta \right) ^{3}}
+\frac{1}{1152} {\frac { \left( 3\,m+1 \right)  \left( m-1 \right)
\left( 3\,m-1 \right) }{ \left( m+1 \right)
^{2}{\alpha}^{2}}} { \left( \alpha-\theta \right) ^{4}}\\
- \frac{1}{172800} {\frac { \left( 3\,m+1 \right)  \left( m-1
\right)  \left( 201\,{m}^{2 }-140\,m+11 \right) }{ \left( m+1
\right) ^{3}{\alpha}^{3}}} { \left( \alpha-\theta
 \right) ^{5}} + \ldots
\end{gathered}
\end{equation}

From the boundary condition (\ref{Gr_lim}) we obtain

\begin{equation}\label{01}
\begin{gathered}
{\frac {{\alpha}^{3} }{2985984000}} \, \frac {\left(
105147\,{m}^{4}+ 384822\,{m}^{3}+519188\,{m}^{2}+307082\,m+66161
\right)} { \left( m+1 \right) ^{6}}\\
{ \left( 9491+
24237\,{m}^{4}+84342\,{m}^{3}+54602\,m+103808\,{m}^{2} \right) } = 1
\end{gathered}
\end{equation}

From equation (\ref{01}) we find

\begin{equation*}
\begin{gathered}
\alpha = 1440\, { \left( m+1 \right)^{2}} / \left(
2548447839\,{m}^{8}+18195239088\,{m}^{7} \right. \\
+55955316456\,{m}^{6} +96920939400\,{m}^{5}+103409323126\,{m}^{4}\\
\left. { +69458768096\,{m}^{3} +28562945760\,{m}^{2} +6527038184\,m
+627934051} \right) ^{\frac{1}{3}}
\end{gathered}
\end{equation*}

\begin{table}[h]
    \center
    \caption{} \label{t:pl_k_n1}
    \begin{tabular}{||c|c|c|c|c|c|c||}     
        \hline
        k & 0 & $ 1 $ & $ 2 $ & $ 3 $ & 4 & 5 \\ \hline
        $ m $ & 1/3 & 1 & 5/3 & 7/3 & 3 & 11/3 \\ \hline
        $ \alpha $ & 1.1762 & 0.7937 & 0.6222 & 0.5211 & 0.4532 & 0.4039 \\ \hline
    \end{tabular}
\end{table}

The values of $ \alpha $, $ m $ and $ k $ at $ n=1 $ are given in
table \ref{t:pl_k_n1}. Approximate solutions of the problem
(\ref{ObshEq_nu}) -- (\ref{Gr_inf}) at $ n=1 $ are expressed by formula (\ref{f3_10}).

For the case $ k=0 $ approximate solution of the boundary value
problem (\ref{ObshEq_nu}) -- (\ref{Gr_inf}) takes the form

\begin{equation}\label{A-sol_p}
f(\theta) = \left\{
\begin{gathered}
\frac{2 \alpha(\alpha- \theta)}{3} -\frac{(\alpha-\theta)^2}{12} +\frac{(\alpha-\theta)^3}{144\alpha}
-\frac{(\alpha-\theta)^5}{23040\alpha^3}, \qquad \\
\hfill 0 < \theta < \alpha;  \\
0, \quad \hfill \alpha < \theta;  \\
\end{gathered}
\right.
\end{equation}

where $ \alpha $ is $ \alpha = 1.1762 $.

\begin{figure}[t] 
\centerline{\epsfig{file=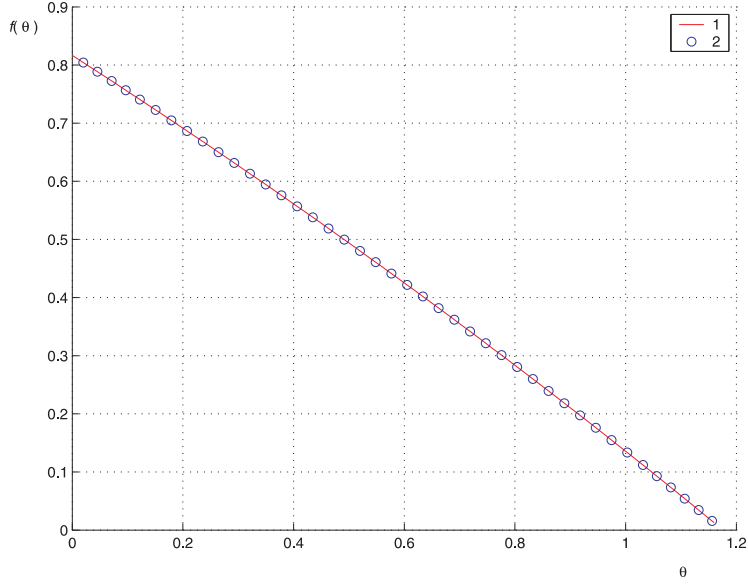,width=100mm}}
\caption{Comparison of approximate solution (\ref{A-sol_p}) and
numerical solution of the problem (\ref{ObshEq_nu}) --
(\ref{Gr_inf}): 1 --- approximate solution; 2 --- numerical
solution.}\label{fig:Zsol_p}
\end{figure}

To check approximate solution (\ref{A-sol_p}) we have compared
it with numerical solution of the boundary value problem (\ref{ObshEq_nu}) -- (\ref{Gr_inf}). The
comparison of approximate solution (\ref{A-sol_p}) and numerical
one at $ k=0 $ and $ n=1 $ is given on Fig.\ref{fig:Zsol_p}. Solid line is approximate solution and circles correspond to numerical
solution of the problem (\ref{ObshEq_nu}) -- (\ref{Gr_inf}). From
Fig.\ref{fig:Zsol_p} we can see that these solutions are similar.

Consider the case $ {n = 4/3} $ and $ k>0 $. Approximate
solution of the boundary value problem (\ref{ObshEq_nu}) --
(\ref{Gr_inf}) is given by formula

\begin{equation}\label{f33_10}
\begin{gathered}
f(\theta) = \frac{2}{3} \alpha \left(m+1 \right) \left(
\alpha-\theta \right) + \frac{1}{7} \left(m-1\right) \left(
\alpha-\theta \right) ^{2 } \\
-{\frac {2}{539}}\,{\frac { \left( 5\,{m}^{2}-2-3\,m \right)  \left(
\alpha-\theta \right) ^{3}}{\alpha\, \left( m+1 \right) }} \\
+{\frac {1}{ 37730}}\,{\frac { \left(
205\,{m}^{3}-188\,{m}^{2}-43\,m+26 \right) \left( \alpha-\theta
\right) ^{4}}{ \left( m+1 \right) ^{2}{\alpha}^{
2}}} \\
-{\frac { 9 \left( 29055\,{m}^{4}+7078\,m+
1199\,{m}^{2}-464-36868\,{m}^{3} \right) \left( \alpha-\theta
\right) ^{5}}{137997475 {\alpha}^{3} \left( m+1 \right) ^{3}}} +
\ldots
\end{gathered}
\end{equation}

Some values of $ \alpha $, $ m $ and $ k $ at $ n= 4/3 $ are given in
table \ref{t:pl_k_n43}.

\begin{table}[h]
    \center
    \caption{} \label{t:pl_k_n43}
    \begin{tabular}{||c|c|c|c|c|c|c||}     
        \hline
        k & 0 & $ 1 $ & $ 2 $ & $ 3 $ & 4 & 5 \\ \hline
        $ m $ & 2/5 & 6/5 & 2 & 14/5 & 18/5 & 22/5 \\ \hline
        $ \alpha $ & 0.9256 & 0.6000 & 0.4608 & 0.3807 & 0.3278 & 0.2897 \\ \hline
    \end{tabular}
\end{table}


Consider the case $ {k = 0} $ and $ n > 0 $. From the expression (\ref{m0k}) we have

\begin{equation*}
m=\frac{n}{n+2}
\end{equation*}

Approximate solutions of the boundary value problem
(\ref{ObshEq_nu}) -- (\ref{Gr_inf}) at $ {k = 0} $, $ n > 0 $ take
the form

\begin{equation}\label{Resh_l1}
\begin{gathered}
f(\theta)=\frac{1}{2}\left( {\frac
{\alpha\,{n}^{2}}{n+2}}+\,\alpha\,n \right) \left( \alpha-\theta
\right)     -\frac{1}{2}\,{\frac {n \left( \alpha-\theta
 \right) ^{2}}{ \left( n+1 \right)  \left( n+2 \right) }}+ \\
+\frac{1}{6}\,{\frac {{n}^{2} \left( \alpha-\theta \right)
^{3}}{\alpha\, \left( n+1  \right) ^{3} \left( 1+2\,n \right) }}
-\frac{1}{24}\,{\frac { \left( 2\,{n}^{2 }+n-3 \right) {n}^{2}
\left( \alpha-\theta \right) ^{4}}{ \left( n+1
 \right) ^{5} \left( 1+2\,n \right) {\alpha}^{2} \left( 3\,n+1
 \right) }}+ \\
+{\frac {1}{120}}\,{\frac {{n}^{2} \left( 12+8\,n-75\,{n}^{
2}+12\,{n}^{5}-77\,{n}^{3} \right)  \left( \alpha-\theta \right)
^{5}} {{\alpha}^{3} \left( n+1 \right) ^{7} \left( 1+2\,n \right)
^{2} \left( 3\,n+1 \right)  \left( 1+4\,n \right) }} + \ldots
\end{gathered}
\end{equation}

Using boundary condition (\ref{Gr_lim}) we obtain the value of $ \alpha $. Some values of $ \alpha $ and
$ n $ at $ k=0 $ are given in table \ref{t:pl_k0}

\begin{table}[h]
    \center
    \caption{} \label{t:pl_k0}
    \begin{tabular}{||c|c|c|c|c|c|c||}     
        \hline
        n & 1 & $ {4}/{3} $ & $ 2 $ & $ {5}/{2} $ & 3 & 4\\ \hline
        $ \alpha $ & 1.4819 & 1.1578 & 0.7889 & 0.6283 & 0.5178 & 0.3775 \\
        \hline  \hline
        n & $ {9}/{2} $ & $ 5 $ & $ {11}/{2} $ & 6 & $ {13}/{2} $ & $ 7 $ \\ \hline
        $ \alpha $ & 0.3307 & 0.2934 & 0.2632 & 0.2382 & 0.2172 & 0.1994  \\ \hline
    \end{tabular}
\end{table}


\section{Conclusion.}

The boundary value problem of the nonlinear heat equation for the
given flow was considered. This problem was solved using the both
numerical and analytical approaches. Some exact solutions were
found. Approximate solutions of the boundary value problem were
obtained. Comparison of the numerical and the approximate solutions
was given.

\bigskip

This work was supported by the International Science and Technology
Center (project B1213).


\bigskip


\end{document}